\documentclass{amsart}
\usepackage{fullpage,amsfonts,amsmath,amscd,amssymb}
\input{xy}
\xyoption{all}
\xyoption{curve}

\usepackage{graphicx}
\usepackage{psfrag}
\usepackage{latexsym}
\usepackage{amsmath}
\usepackage{amscd}
\usepackage{amssymb}
\usepackage{xspace}
\usepackage{theorem}
\usepackage{multirow}

\newcommand{\F}{{\mathbb{F}}}
\newcommand{\tF}{{\widetilde{\mathbb{F}}}}
\newcommand{\hF}{{\widehat{\mathbb{F}}}}

\newcommand{\bea}{\begin{eqnarray}}
\newcommand{\eea}{\end{eqnarray}}
\usepackage{amsmath}
\newenvironment{ea}{ \begin{equation}\begin{aligned} }{ \end{aligned}\end{equation} }
\newcommand{\xor}{\oplus}

\newenvironment{mylist}{\begin{list}{}{
\setlength{\itemsep}{0mm}
\setlength{\parskip}{0mm}
\setlength{\topsep}{1mm}
\setlength{\parsep}{0mm}
\setlength{\itemsep}{0mm}
\setlength{\labelwidth}{6mm}
\setlength{\labelsep}{3mm}
\setlength{\itemindent}{0mm}
\setlength{\leftmargin}{9mm}
\setlength{\listparindent}{6mm}
}}{\end{list}}

\begin{document}
%
\title{An approach to RAID-6 based on cyclic groups} 
\author{Robert Jackson}
\address{Arithmatica, ltd., Haseley Business Centre, Warwick, CV35 7LS, UK}
\email{robert.jackson@arithmatica.com}
\author{Dmitriy Rumynin}
\address{Department of Mathematics, University of Warwick, Coventry, CV4 7AL, UK}
\email{D.Rumynin@warwick.ac.uk}
\author{Oleg V. Zaboronski}
\address{Department of Mathematics, University of Warwick, Coventry, CV4 7AL, UK}
\email{O.V.Zaboronski@warwick.ac.uk}
\thanks{Oleg V. Zaboronski was partially supported by Royal Society Industrial Fellowship}
\date{June 3, 2008, revised on March 26, 2010}
\subjclass{Primary 94B60; Secondary 11A41}
\keywords{RAID-6, Artin's conjecture, Mersenne prime, Reed-Solomon Code} 

\begin{abstract}
As the size of data storing arrays of disks grows, it becomes vital to protect
data against double disk failures.
A popular method of protection is via 
the Reed-Solomon (RS) code with two parity words.
In the present 
paper we construct alternative
examples of linear block codes protecting against two erasures.
Our construction is based on an abstract notion of cone.
Concrete cones are constructed via matrix representations of cyclic groups of prime
order. In particular, this construction produces EVENODD code. 
Interesting conditions on the prime number arise 
in our analysis of these codes. 
At the end, we analyse an
assembly implementation of the corresponding system on a general
purpose processor and 
compare its write and recovery
speed with the standard DP-RAID system.
\end{abstract}

\maketitle


\section{Introduction}
%
%
%


A typical storage solution targeting a small-to-medium size enterprise is a networked unit 
with 12 disk drives with total capacity of around 20 TB \cite{net, rei}.
The volume of information accumulated and stored by a typical small-size
information technology company amounts to fifty 100-gigabyte drives.
The specified mean time between failures (MTBF) for a modern desktop
drive is about 500,000 hours \cite{fuj}. Assuming that such an MTBF is
actually achieved and that the drives fail independently, the
probability of a disk failure in the course of a year is
$1-e^{-12/57}\approx 0.2$. Therefore, even a small company can no longer avoid the necessity of protecting
its data against disk failures. The use of redundant arrays of independent disks (RAIDs) enables such a protection
in a cost efficient manner.

To protect an array of $N$ disks against a single disk failure
it is sufficient to add one more disk to the array. For every $N$ bits of
user data written on $N$ disks of the array, a parity bit equal to an exclusive OR (XOR) of these bits
is written on the $(N+1)$-st disk. Binary content of any disk can be then recovered as a bitwise XOR of contents
of remaining $N$ disks. The corresponding system for storing data and
distributing parity between disks of the array is referred to
as RAID-5 \cite{bob}. Today, RAID-5 constitutes the most popular solution for protected storage.

As the amount of data stored by humanity  on magnetic media grows, the danger of $multiple$ disk failures
within a single array becomes real. Maddock, Hart and Kean argue
that for a storage system consisting of one hundred $8+P$ RAID-5 arrays
the rate of failures amounts to losing one array every six months \cite{bob}.
Because of this danger, RAID-5 is currently being replaced with RAID-6, which offers protection against
double failure of drives within the array. RAID-6 refers to any technique where
two strips of redundant data are added to the strips of user data, in such a way that all the information
can be restored if any two strips are lost.

A number of RAID-6 techniques are known \cite{hel,bob,sri}.
A well-known RAID-6 scheme is based on the rate-$255/257$ Reed-Solomon code \cite{peter}.
In this scheme two extra disks are introduced for up to 255 disks of data and two parity bytes are computed per
255 data bytes. Hardware implementation of RS-based RAID-6 is as simple as operations in $\tF=GF(256)$, which are byte-based. Addition of
bytes is just a bitwise XOR. Multiplication of bytes
corresponds to multiplication of boolean polynomials modulo an irreducible polynomial. Multiplication
can be implemented using XOR-gates, AND-gates and shifts.

Some RAID-6 schemes use only bitwise XOR  for the computation of parity bits
by exploiting a two-dimensional striping of disks of the array.
Examples are a proprietary RAID-DP developed by Network Appliances \cite{na}
and EVENODD \cite{bbbm}.
Some other RAID-6 methods use a non-trivial striping and employ only
XOR operation for parity calculation and reconstruction. Examples 
include X-code, ZZS-code and Park-code \cite{bob,xb,zzs}.

In all the cases mentioned above, the problem dealt with is inventing
an error correcting block code capable of correcting up to two erasures (we assume that
it is always known which disks have failed). In the present paper we describe a general approach to the solution
of this problem, which allows one to develop an optimal RAID-6 scheme for given
technological constraints (e.g. available hardware, the number of disks in the array, the required read
and write performance). We also consider an assembly implementation of an exemplary RAID-6 system built using our
method and show that it outperforms the Linux kernel implementation of RS-based RAID-6.

The paper is organised as follows. In Section 2 we discuss RAID-6 in the context of systematic linear block codes
and construct simple examples of codes capable of correcting two errors in known positions. In Section 3 we identify
an algebraic structure (cone) 
common to all such codes and use it to construct RAID-6 schemes starting with elements
of a cyclic group of a prime order. 
Section 3.3 is of particular interest to number theorists where we discuss a new condition on the prime numbers arising in
the context of RAID-6 schemes.
In Section 4 we compare encoding and decoding performances of an assembly implementation of RAID-6 based on
$\mathbf{}{Z}_{17}$ with its RS-based counterpart implemented as a part of Linux kernel.

Let us comment on the relation of the presented material to other modern research efforts.
Section 2 is rather standard  \cite{bla}. All original theoretical material of this paper is in Section 3. 
The notion of a cone is somewhat related to a non-singular difference set of Blaum and Roth \cite{bla}
but there are essential differences between them. The cone from a cyclic group of prime order
as in Lemma~\ref{26}.1 gives EVENODD code \cite{bbbm}. Its extended versions and connections to number theoretic conditions
are new.

\section{RAID-6 from the viewpoint of linear block codes.}
Suppose that information to be written on the array of disks is broken into
words of length $n$ bits. What is the best rate linear block code,
which can protect data against the loss of two words?

Altogether, there are $2^{2n}$ possible pairs of words. In order
to distinguish between them, one needs at least $2n$ distinct syndromes.
Therefore, any linear block code capable of restoring $2$
lost words in known locations must have at least $2n$ parity
checks. Suppose the size of the information block is $Nn$ bits or $N$ words.
In the context of RAID, $N$ is the number of information disks to
be protected against the failure. Then the code's block size must
be at least $(2+N)n$ and the rate is
$$
R\leq \frac{N}{N+2}
$$
This result is intuitively clear: to protect $N$ information disks
against double failure, we need at least $2$ parity disks. Note
however, that in order to achieve this optimal rate, the word
length $n$ must grow with the number of disks $N$. Really, the
size of parity check matrix is $2n \times (N+2)n$. All columns of
the matrix must be distinct and nonzero. Therefore, $(N+2)n \leq
(2^{2n}-1)$, i. e.
$$
\frac{1}{n}\bigg( 2^{2n}-1-2n \bigg)\geq N
$$
If in particular $n=1$, then $N\leq 1$. Therefore, if one wants to
protect information written on the disks against double failures
using just two parity disks, the word size $n\geq 2$ is necessary.
If $n=2$, we get $N\leq 5$. In reality, the lower bound on $n$
(or, equivalently, the upper bound on $N$) is more severe, as the
condition that all columns of parity check matrix are distinct
leads to a code with minimal distance $d_{min}=2$. However, in
order to build a code which corrects up to $2n$ errors in the
known location we need $d_{min}=2n$.

In the following subsections we will construct explicit examples of
linear codes for RAID-6 for small values of $n$ and $N$. These examples both guide
and illustrate our general construction of RAID-6 codes presented in Section 3.

\subsection{Redundant array of four independent disks, which
protects against the failure of any two disks.\label{21}}

We restrict our attention to {\em systematic} linear block codes.
These are determined by the parity matrix. To preserve a
backward compatibility with RAID-5 schemes, we require half of the
parity bits to be the straight XOR of the information bits. Hence
the general form of the parity check matrix for $N=2$ is
 \bea
 P=\bigg(\begin{array}{cccc}
 I_{n\times n} & I_{n\times n} & I_{n\times n}& 0_{n\times n} \\
 H & G & 0_{n\times n} & I_{n\times n},
 \end{array}\bigg)\label{parity}
 \eea
 where $I_{n\times n}$ and $0_{n\times n}$ are $n\times n$
 identity and zero matrix correspondingly; $G$ and $H$ are
 some $n\times n$ binary matrices.
The corresponding parity check equations are
\begin{ea}
d_{1}+d_{2}+\pi_{1}=0
\hspace{10mm} \hbox{ and } \hspace{10mm}
H\cdot d_{1}+G\cdot
d_{2}+\pi_{2}=0.\label{pp2}
\end{ea}
Here $d_{1}, d_{2}$ are $n$-bit words written on disks 1 and 2,
$\pi_{1}$ and $\pi_{2}$ are $n$-bit parity check words written on
disks 3 and 4; "$\cdot$" stands for binary matrix multiplication.

Matrices $G$ and $H$ defining the code are constrained by the
condition that the system of parity check equations must have a
unique solution with  respect to $any$ pair of variables. To determine these constraints we
need to consider the following particular cases. 
\begin{mylist}
\item[{\em $(\pi_{1}, \pi_{2})$ are lost.}] The system (
\ref{pp2}) always has a unique solution with respect to lost variables: we can compute
parity bits in terms of information bits.
\item[{\em $(d_{1}, \pi_{2})$ are lost.}] The system (
\ref{pp2})
always has a unique solution with respect to lost variables: compute $d_{1}$ in terms of
$\pi_{1}$ and $d_{2}$ using the first equation of (\ref{pp2}) as in RAID-5.
Then compute $\pi_{2}$ using the second equation. 
\item[{\em $(d_{2}, \pi_{2})$ are lost.}] The system (\ref{pp2}) always has a unique
solution with respect to lost variables: compute $d_{2}$ using $\pi_{1}$ and $d_{1}$ as in
RAID-5.
Then compute $\pi_{1}$ using (\ref{pp2}).
\item[{\em $(\pi_{1}, d_{1})$ are lost.}] The system (\ref{pp2}) always has a unique
solution with respect to lost variables
provided the matrix $H$ is invertible.
\item[{\em $(\pi_{1}, d_{2})$ are lost.}] The system (\ref{pp2}) always has a unique
solution with respect to lost variables
provided the matrix $G$ is invertible.
\item[{\em $(d_{1}, d_{2})$ are lost. }]
The system (\ref{pp2}) always has a unique
solution with respect to lost variables
provided the matrix 
$ \bigg( \begin{array}{cc}
 I_{n\times n} & I_{n\times n}\\
 H_{n\times n} & G_{n\times n}
 \end{array} \bigg)
 \label{nondeg}$
is invertible.
\end{mylist}
As it turns out, one can build a parity check matrix satisfying
all the non-degeneracy requirements listed above for $n=2$. The
simplest choice is
 \bea
 H=I_{2\times 2},~G=\bigg( \begin{array}{cc}
 0 & 1\\
 1 & 1
 \end{array} \bigg).\label{hg}
 \eea
Non-degeneracy of the three matrices $H$, $G$ and (\ref{nondeg}) is evident. 
For instance\footnote{The reader is aware that $-1=1\neq 0$ in characteristic 2},
$$
\mbox{det}
 \bigg( \begin{array}{cc}
 I_{n\times n} & I_{n\times n}\\
 H_{n\times n} & G_{n\times n}
 \end{array} \bigg)
=-1=1.
$$


We conclude that the linear block code with a $4\times 8$ parity
check matrix (\ref{parity}, \ref{hg}) gives rise to RAID-6
consisting of four disks. The computation of parity dibits
$\pi_{1},~\pi_{2}$ in the described DP RAID is almost as simple as
the computation of regular parity bits: Let
$d_{1}=(d_{11},d_{12})$ and $d_{2}=(d_{21},d_{22})$ be the dibits
to be written on disks one and two correspondingly. Then
\begin{eqnarray}
 \pi_{11}=d_{11}+d_{21},&\hspace{40mm}&
 \pi_{12}=d_{12}+d_{22},\nonumber\\
 \pi_{21}=d_{11}+d_{22},&\hspace{40mm}&
 \pi_{22}=d_{12}+d_{21}+d_{22}.\label{hordiag}
 \end{eqnarray}
The computations involved in the recovery of lost data are
 bitwise XOR only. As an illustration, let us write down
 expressions for lost data bits in terms of parity bits
 explicitly:
 \bea
 d_{22}=\pi_{11}+\pi_{12}+\pi_{21}+\pi_{22}, &\hspace{20mm}
 &d_{12}=\pi_{11}+\pi_{21}+\pi_{22}, \nonumber\\
 d_{11}=\pi_{11}+\pi_{12}+\pi_{22},\hspace{10mm} &\hspace{30mm}
 &d_{21}=\pi_{12}+\pi_{22}.\nonumber
 \eea
It is interesting to note that RAID-6 code described here
is equivalent to Network Appliances' horizontal-diagonal parity RAID-DP$^{TM}$
with two data disks \cite{na}. Really, diagonal-horizontal parity
system for two info disks is
\begin{displaymath}
\begin{array}{cccc}
A&B&HP&DP1\\
C&D&HP2&DP2,
\end{array}
\end{displaymath}
where strings $(A,C)$ are written on information disk 1, strings
$(B,D)$ are written on disk 2, $(HP, HP2)$ is horizontal parity,
$(DP1, DP2)$ is diagonal parity. By definition, $HP=A+B$,
$HP2=C+D$, $DP1=A+D$, $DP2=B+C+D$, which coincides with parity
check equations (\ref{hordiag}).

On the other hand, the code (\ref{parity}) is a reduction of the RS code based on $GF(4)$
which we will describe in the next subsection.

\subsection{Redundant array of five independent disks, which
protects against the failure of any two disks.\label{22}}

The code (\ref{parity}) can
be extended to a scheme providing double protection of user data
written on three disks \cite[Example 1.1]{bla}. The parity check matrix is
 \bea
 P=\bigg(\begin{array}{ccccc}
 I_{2\times 2} & I_{2\times 2} &I_{2\times 2}& I_{2\times 2}& 0_{2\times 2} \\
 I_{2\times 2} & G & G^2 & 0_{2\times 2} & I_{2\times 2},\label{gf4}
 \end{array}\bigg)
 \eea
where $2\times 2$ matrix $G$ was defined in the previous
subsection. The corresponding parity check equations are
 \begin{equation}
 d_{1}+d_{2}+d_{3}+\pi_{1}=0, \hspace{20mm}
 d_{1}+G\cdot d_{2}+G^2\cdot d_{3}+\pi_{2}=0.\label{p52}
 \end{equation}

 The solubility of these equations with respect to any pair of
 variables from the set $\{d_{1}, d_{2}, d_{3}, \pi_{1}, \pi_{2}
 \}$ requires two extra conditions of non-degeneracy in addition to
 non-degeneracy conditions listed in the previous subsection. Namely,
 matrices
 $
 \bigg( \begin{array}{cc}
 I_{2\times 2} & I_{2\times 2}\\
 I_{2\times 2} & G^2
 \end{array} \bigg)$
 and
$ \bigg( \begin{array}{cc}
 I_{2\times 2} & I_{2\times 2}\\
 G & G^2
 \end{array} \bigg)$
must be invertible. It is possible to check the invertibility of
these matrices via a direct computation. However, in the next
section we will construct a generalisation of the above example
and find an elegant way of proving non-degeneracy.

The code (\ref{parity}) is a reduction of (\ref{gf4}) corresponding to $d_3=0$. Note also that
the code (\ref{gf4}) is equivalent to rate-$3/5$ Reed-Solomon code based on $GF(4)$:
a direct check shows that the set of $2\times 2$ matrices $0,1,G,G^2$ is closed under
multiplication and addition and all non-zero matrices are
invertible. Thus this set forms a field isomorphic to $GF(4)$.
On the other hand, as we established in the previous subsection,
the code (\ref{parity}) is equivalent to RAID-DP$^{TM}$ with four disks.
Therefore, RAID-DP$^{TM}$ with four disks is a particular case of the RS-based RAID-6.
It would be interesting to see if RAID-DP$^{TM}$ can be reduced to the RS-based RAID-6 in general.

We are now ready to formulate general properties of linear block codes suitable for RAID-6 and construct
a new class of such codes.

\section{RAID-6 based on the cyclic group of a prime order.}
\subsection{RAID-6 and cones of $GL_{n}(\F)$.\label{23}}
In this subsection we will define a general mathematical object
underlying all existing algebraic RAID-6 schemes. 
We recall that $\F=GF(2)$ is the field of two elements
and $GL_n(\F)$ is the set of $n\times n$ invertible matrices. 

{\bf Definition \ref{23}.1.} A cone\footnote{
This terminology is slightly questionable. If one asks $g+h \in C$
then $C\cup\{ 0\}$ is a convex cone in the usual mathematical sense.
Our choice of the term is influenced by this analogy.
Non-singular difference set or quasicone or RAID-cone could be more appropriate scientifically
but would pay a heavy linguistic toll.
}
$C$ is a subset of
$GL_{n}(\F)$ such that 
$g+h \in GL_n(\F)$
for all $g \neq h \in C$. 

This notion is related to {\em non-singular difference sets} of Blaum and Roth \cite{bla}. 
The cone satisfies the axioms P1 and P2 of Blaum and Roth but the final axiom P3 or P3' is too restrictive
for our ends. 
On the other hand, we consider only binary codes while Blaum and Roth consider codes over any finite field.

A standard example of a cone appears in the context of Galois fields. If we choose a basis of $GF(2^n)$ as a vector space over $\F$
then we can think of $GL_{n}(\F)$ as the group of all $\F$-linear transformations of $GF(2^n)$. Multiplications by non-zero elements
of $GF(2^n)$ form a cone. If $\alpha\in GF(2^m)$ is a primitive generator and $g$ is the matrix of multiplication by  $\alpha$ 
then this cone is $\{ g^m \; | \; 0\leq m\leq 2^n-2 \}$.
This cone gives the RS-code with two parity words. 

The usefulness of cones for RAID-6 is explained by the
following

{\bf Lemma \ref{23}.2.} {\it Let $C = \{g_1,g_2,\ldots g_N \} \subseteq
GL_n(\F)$ be a cone of $N$ elements. Then the system of parity equations
 \bea
 d_{N+1}=\sum_{k=1}^N d_{k} \hspace{20mm} and \hspace{20mm}
 d_{N+2}=\sum_{k=1}^N g_{k}d_{k}\label{cone}
 \eea
has a unique solution with respect to any pair of variables $(d_i, d_j)
\in \F^n\times \F^n $, $1\leq i<j\leq N+2$. Here $d_{i}$ are binary
$n$-dimensional vectors.}

{\bf Proof.} The fact that system (\ref{cone}) has a unique
solution with respect to $(d_{N+1},d_{N+2})$ is obvious.

The system has a unique solution with respect to $(d_{N+1},d_{j})$ for
any $j\leq N$: from the second of equations (\ref{cone}),
$d_j=g_{j}^{-1}(d_{N+2}+\sum_{k\neq j}^N g_k d_k)$, where we used
invertibility of $g_j \in GL_n(\F)$. With $d_j$ known, $d_{N+1}$ can be
computed from the first of equations (\ref{cone}).

The system has a unique solution with respect to $(d_{N+2},d_{j})$ for
any $j\leq N$: from the first of equations (\ref{cone}),
$d_j=d_{N+1}+\sum_{k\neq j}^N d_k$. With $d_j$ known, $d_{N+2}$
can be computed from the second of equations (\ref{cone}).

The system has a unique solution with respect to any pair of variables
$d_i,d_j$ for $1\leq i<j\leq N$: multiplying the first of
equations (\ref{cone}) with $g_i$ and adding the first and second
equations, we get $d_j=(g_i+g_j)^{-1}(g_i d_{N+1}+d_{N+2}+\sum_{k\neq
i,j}^{N} (g_k+g_i) d_k)$. Here we used the invertibility of the sum $g_i+g_j$
for any $i\neq j$, which follows from the definition of the cone.
With $d_j$ known, $d_i$ can be determined from
any of the equations (\ref{cone}). {\bf QED}

In the context of RAID-6, $d_i$ for $1\leq i\leq N$ can be
thought of as $n$-bit strings of user data, $d_{N+1}, d_{N+2}$ - as $n$-bit
parity strings. The lemma proved above ensures that any two strings
can be restored from the remaining $N$ strings.

We conclude that any cone can be used to build RAID-6.
The following lemma 
gives some necessary conditions for a cone. 

{\bf Lemma \ref{23}.3.} {\it Let $C\subset GL_n(\F)$ be a cone.
\begin{mylist}
\item[(i)] For all $ g,~h \in C$ such that $g \neq h$ and for all $x \in GF(2^m)^n$,
$gx=hx$ if and only if $x=0$. 
\item[(ii)] No two elements of
the same cone can share an eigenvector in $\F^n$. 
\item[(iii)] The cone $C$
can contain no more than one permutation
matrix.
\end{mylist}
}

{\bf Proof.} To prove (i), assume that there is $x \neq 0:$ $gx=hx$. Then
$(g+h)x=0$, which contradicts the fact that $g+h$ is
non-degenerate. Therefore, $x=0$. Let us prove (ii) now.
As elements of $C$ are non-degenerate, the only
possible eigenvalue  in $\F$ is $1$, thus for
any two elements sharing an eigenvector $x$, $x=hx=gx$, which
again would imply degeneracy of $h+g$ unless $x=0$. The statement (iii) follows from
(ii) if one notices that any two permutation matrices share an
eigenvector whose components are all equal to one. {\bf QED}

The notion of the cone is convenient for restating well understood
conditions for a linear block code to be capable of recovering up to two lost words.
Our main challenge is to find examples of cones with sufficiently many
elements, which lead to easily implementable RAID-6 systems.
We will now construct a class of cones starting
with elements of a cyclic subgroup of $GL_n(\F)$  of a prime order.

\subsection{RAID-6 based on matrix generators of $Z_{N}$.\label{24}}
We start with the following

{\bf Theorem \ref{24}.1.} {\it Let $N$ be an odd number. Let
$g$ be an $n \times n$ binary matrix 
such that 
$g^{N}=Id$
and 
$Id+g^m$ is
non-degenerate for each proper\footnote{a natural number $m<N$ that divides $N$} divisor $m$ of $N$. 
Then the elements of cyclic
group $Z_N=\{ Id, g, g^2, \ldots, g^{N-1} \}$ form a cone.}

The proof of the Theorem \ref{24}.1 is based on the following two lemmas.

 {\bf Lemma \ref{24}.2.} {\it Let $g$ be a binary matrix such
that $Id+g$ is non-degenerate and $g^N=Id$, where $N$ is an
integer. Then
 \bea
 \sum_{k=0}^{N-1}g^k=0\label{rootsum}
 \eea
 }

 {\bf Proof.} Let us multiply the left hand side of (\ref{rootsum}) with $(Id+g)$ and
 simplify the result using that $h+h=0$ for any binary matrix:
\begin{eqnarray*}
 (Id+g)\sum_{k=0}^{N-1}g^k=Id+g+g+g^2+\ldots
+ g^{N-1}+g^{N-1}+g^{N}=
Id+g^N=Id+Id=0.
\end{eqnarray*}
As $Id+g$ is non-degenerate, this implies that
$\sum_{k=0}^{N-1}g^k=0$. {\bf QED}

Lemma \ref{24}.2 is a counterpart of a well-known fact from
complex analysis that roots of unity add to zero.

 {\bf Lemma \ref{24}.3.} {\it Let $g$ be a binary matrix such that
$g^N=Id$ for an odd number $N$
and
$Id+g^m$ is non-degenerate for every proper divisor $m$ of $N$.
Then the matrix $g^l+g^k$ is non-degenerate for any
$k,l:~0\leq k<l<N$.}

{\bf Proof.} As $g^N=Id$, the matrix $g$ is invertible. To prove
the lemma, it is therefore sufficient to check the non-degeneracy of
$Id+g^k$ for $0<k<N$.

The group $Z_{N}=\{1,g,g^2,\ldots, g^{N-1} \}$ is cyclic.
An element $g_{k}=g^k$ for $0<k<N$ generates the cyclic subgroup $Z_{N/d}$
where $d$ is the the greatest common divisor of $N$ and $k$ and the element
$g^d$ generates the same subgroup.
Since the matrix $g^d$ satisfies all the conditions of Lemma
\ref{24}.$2$, the sum of all elements of $Z_{N/d}$ is zero.
Therefore,
 \bea 0=\sum_{m=0}^{\frac{N}{d}-1}
g_{k}^m=(Id+g_k)+g_k^2(Id+g_k)+\ldots 
+g_k^{N-3}(Id+g_k)+g_k^{N-1}=0.\label{242}
 \eea
The grouping of terms used in (\ref{242}) is possible as $N/d$ is
odd. Assume that matrix $1+g_{k}$ is degenerate. Then there exists
a non-zero binary vector $x$ such that $(1+g_{k})x=0$. Applying
both sides of (\ref{242}) to $x$ we get $g_k^{N-1} x=g^{k(N-1)}
x=0$. This contradicts non-degeneracy of $g$. Thus the
non-degeneracy of
$1+g^k$ is proved for all $0<k<N$. {\bf QED}

{\bf The proof of Theorem \ref{24}.1.} The matrix $g$ described in the
statement of the theorem satisfies all requirements of Lemma
\ref{24}.3. The statement of the theorem follows from Definition
\ref{23}.1 of the cone. {\bf QED}

Theorem \ref{24}.1 allows one to determine whether elements of
$Z_{N}$ belong to the same cone by verifying a single
non-degeneracy conditions imposed on the generator.

The following corollary of Theorem \ref{24}.1 makes an explicit link between the constructed cone
and  RAID-6:

 {\bf Corollary \ref{24}.4.} {\it Let $g$ be an $n \times n$ binary
matrix such that
$g^N=Id$ for an odd number $N$
and
$Id+g^m$ is non-degenerate for every proper divisor $m$ of $N$.
The systematic linear block code defined  by the
parity check matrix
 \begin{displaymath}
 P=\bigg(\begin{array}{ccccccc}
 I_{n\times n} & I_{n\times n} &I_{n\times n} &\ldots &I& I& 0 \\
 I_{n\times n}  & g  &g^2 & \ldots&g^{N-1} & 0& I_{n\times n} ,
 \end{array}\bigg)
 \end{displaymath}
can recover up to $2$ n-bit lost words in known positions.
Equivalently, the system of the parity check equations
 \bea
 d_1+d_{2}+\ldots +d_{N}+d_{N+1}=0 \nonumber \\
 d_{1}+gd_{2}+\ldots+g^{N-1}d_{N}+d_{N+2}=0\label{znparity}
 \eea
has a unique solution with respect to any pair of variables $(d_i,d_j)$,
$1\leq i<j\leq N+2$.}

{\bf Proof.} It follows from Theorem \ref{24}.1. that the first $N$
powers of $g$ belong to a cone. The statement of the corollary is
an immediate consequence of Lemma \ref{23}.2 for $g_k=g^{k-1}$, $1\leq k \leq N$. {\bf QED.}

As a simple application of Theorem \ref{24}.1, let us show that
the parity check matrix (\ref{gf4}) does indeed satisfy all
non-degeneracy requirements. The matrix
$ G=\bigg( \begin{array}{cc}
 0 & 1\\
 1 & 1
 \end{array} \bigg)$
is non degenerate and has order $3$. Also, the matrix
$ Id+G=\bigg( \begin{array}{cc}
 1 & 1\\
 1 & 0
 \end{array} \bigg)$
is non-degenerate. Hence in virtue of Corollary \ref{24}.4, the parity
check matrix (\ref{gf4}) determines a RAID system
consisting of five disks, that protects against the failure of any
two disks.
 
\subsection{Extension of $Z_N$-based cones for certain primes. \label{25}} 


 We will now show that for certain primes, the cone constructed in the previous
 subsection can be extended. The existence of such extensions give some curious conditions on a prime number, one of which is new
to the best of our knowledge.
 We start with the following

{\bf Lemma \ref{25}.1.}  {\it Let $N>2$ be a prime number.
Then the group ring $R=\F Z_N$ of the cyclic group of order $N$ is isomorphic to
$\F \oplus \hF^k$ where $\hF = GF (2^d)$, $d$ is the smallest positive integer such that $2^d = 1 \mod N$, $k=(N-1)/d$.}

{\bf Proof.} 
By the Chinese Remainder Theorem, $R\cong \oplus_{j=0}^k \F[X]/(f_j)$ where $X^N-1=f_0\cdot f_1 \cdots f_k$ is the decomposition 
into irreducible over $\F$ polynomials
and $f_0 = X-1$. 
Let $\alpha$ be a root of $f_j$ for some $j>0$. Then $d=\deg f_j$ is the smallest number such that $\alpha\in GF (2^d)\cong \F [X]/(f_j)$.
Hence, $\alpha^{2^d-1}=1$ and $d$ is the smallest with such property.
As $N$ is prime, $\alpha$ is a primitive $N$-th root of unity. Hence, $N$ divides $2^d-1$ and $d$ is the smallest with such property.

It follows that for $j>0$ all $f_j$ have degree $d$ and all   $\F[X]/(f_j)$ are isomorphic to  $GF (2^d)$.
{\bf QED}.

One case of particular interest is $k=1$ which happens when $2$ is a primitive $N-1$-th root of unity modulo $N$.
This forces $k=1$ and $d=N-1$.
Such primes in the first hundred are 3, 5, 11, 13, 19, 29, 37, 53, 59, 61, 67, 83 \cite{seq}.
If a generalised Riemann's hypothesis holds true then there are infinitely many such primes \cite{hoo}.

{\bf Lemma \ref{25}.2.}  {\it Let $N>2$ be a prime number such that  $2$ is a primitive $N-1$-th root of unity modulo $N$.
Let $g$ be an $n \times n$ binary matrix such that $Id+g$ is
non-degenerate and $g^{N}=Id$. Then 
the set of $2^{N-1}-1$ matrices
$S = \{ g^{a_1}+g^{a_2}+ \ldots + g^{a_t} \ \mid\ 0\leq a_1 < a_2 \ldots < a_t < N, \ 1\leq t \leq \frac{N-1}{2}$ is a cone.\} }\\

{\bf Proof.} Matrix $g$ defines a ring homomorphism $\phi:R \rightarrow M_n(\F), \ \phi (\sum_k \alpha_k X^k) = \sum_k \alpha_k g^k$
from the group ring to a matrix ring.
Since $1+g$ is invertible,
$$
1+g+g^2+\ldots+g^{N-1} = (1+g^N)(1+g)^{-1}=0
$$
and $1+X+\ldots+X^{N-1}$ lies in the kernel of $\phi$.
Since  $R/(1+X+\ldots+X^{N-1})\cong\F[X]/(f_1)\cong \hF$, the image $\phi (R)$ is a field and $S$ is a subset of $\phi (R)$.
Finally, as $f_1=1+X+\ldots+X^{N-1}$ is the minimal polynomial of $g$, all elements  $g^{a_1}+g^{a_2}+ \ldots + g^{a_t}$ listed above
are distinct and nonzero and $S=\phi (R) \setminus \{ 0 \}$.
{\bf QED}.

Notice that for $N=3$, the set $S$ consists only of $Id$ and $g$.

The cone $S$ in Lemma \ref{25}.2. may be difficult to use in a real system but it contains a very convenient subcone as soon as $N>3$. This
subcone consists of elements $g^i$ and $Id+g^j$. The following theorem gives a condition on the prime $p$ for these elements to form a cone.
This condition is new to the best of our knowledge.

 {\bf Theorem \ref{25}.3.}  {\it  
The following conditions are equivalent for a prime number $N>2$.
\begin{mylist}
\item[(1)] For any $n \times n$ binary matrix $g$ such that $Id+g$ is
non-degenerate and $g^{N}=Id$ the set of $2N-1$ matrices
$S = \{ Id, g, g^2, \ldots, g^{N-1}, Id+g, Id+g^2,\ldots, Id+g^{N-1} \}$ is a cone.
\item[(2)] 
For no primitive $N$-th root of unity $\alpha$ in the algebraic closure of $\F$, the element $\alpha+1$ is an $N$-th root of unity.
\item[(3)] For any $0<m<N$ the polynomials $X^N+1$ and $X^m+X+1$ 
are relatively prime.
\item[(4)] No primitive $N$-th root of unity $\alpha$ in the algebraic closure of $\F$ satisfies  $\alpha^m+\alpha^l+1 = 0$ with $N>m>l>0$.
\end{mylist}
}

{\bf Proof.} 
First, we observe that (1) is equivalent to (4).
If (4) fails, there exists an $N$-th root of unity $\alpha$ such that $\alpha^m+\alpha^l+1 = 0$.
Let $f(X)$ be the minimal polynomial of $\alpha$.
The matrix $g$ of multiplication by the coset of $X$ in
$\F[X]/(f)$ fails condition (1) with $g^m+g^l+1= 0$.

If (4) holds and $g$ is a matrix as in (1) then
the elements of $S$ are all invertible matrices by Theorem \ref{24}.1.
Moreover, it only remains to establish that 
each matrix $g^m+g^l+Id$, $N>m>l>0$ is invertible.
Suppose that it is not invertible. It must have an eigenvector $v\in\F^n$ with the zero eigenvalue.
It follows that 
$f_v(X)$, the minimal polynomial of $g$ with respect to $v$, divides both $X^N+1$ and $X^m+X^l+1$.
Since 1 is not a root of $X^m+X^l+1$, any root $\alpha$ of $f_v(X)$ in the algebraic closure of $\F$
is a primitive $N$-th root of unity and  satisfies  $\alpha^m+\alpha^l+1 = 0$.

Equivalence of (4) and (3) is clear: $\beta=\alpha^l$ is also a primitive root, hence condition (4)
can be rewritten as no root $\beta$ satisfies $\beta^s + \beta +1=0$ with $N>s>0$.
Thus, $X^N+1$ and $X^m+X+1$ do not have common roots in the algebraic closure of $\F$ and must be relatively prime.

Equivalence of (3) and (2) comes from rewriting $\alpha^m + \alpha +1=0$  as  $\alpha^m = \alpha +1$ and observing
that $\alpha^m$ is necessarily a primitive $N$-th root of unity.
{\bf QED}.

This theorem allows us to sort out whether any particular prime $N$ is suitable for extending the cone.

{\bf Corollary \ref{25}.4.}  {\it A Fermat prime $N>3$ satisfies the conditions of Theorem \ref{25}.3.
A Mersenne prime fails the conditions of Theorem \ref{25}.3.}

{\bf Proof.} 
A Fermat prime is of the form $N=2^k+1$. Hence, for a primitive $N$th root of unity $\alpha$
$$
(\alpha +1)^N= (\alpha +1)^{2^k}(\alpha +1)= (\alpha^{2^k}+1)(\alpha +1) 
= \alpha^{-1}+\alpha.
$$
If this is equal 1, then $\alpha^{2}+\alpha+1 =0$, forcing $N=3$. 
A Mersenne prime is of the form $N=2^k-1$. Hence, 
$$
(\alpha +1)^N= (\alpha +1)^{2^k}(\alpha +1)^{-1}= (\alpha^{2^k}+1)(\alpha +1)^{-1}=(\alpha+1)(\alpha +1)^{-1}= 1.
$$
{\bf QED}.

In fact, most of the primes appear to satisfy the conditions of Theorem~\ref{25}.3.
In the first 500 primes, the only primes that fail are Mersenne and 73.
Samir Siksek has found several more primes that fail but are not Mersenne. These are 
(in the bracket we state the order of 2 in the multiplicative group of $GF(p)$) 
73 (9)
178481 (23),
262657 (27),
599479 (33),
616318177 (37),
121369 (39),
164511353 (41),
4432676798593 (49),
3203431780337 (59),
145295143558111 (65),
761838257287 (67),
10052678938039 (69),
9361973132609 (73),
581283643249112959 (77).
It would be interesting to know whether there are infinitely many primes failing the conditions of Theorem \ref{25}.3.

Utilising the cone in Theorem \ref{25}.3., we start with a matrix
generator of the cyclic group of an appropriate prime order $N$ to
build a RAID-6 system protecting up to $2N-1$ information disks.
The explicit expression for Q-parity is
 \bea
 Q=\sum_{k=0}^{N-1} g^k d_k+\sum_{k=N}^{2N-2} (Id+g^{k-N+1})d_k,
 \label{qext}
 \eea
 where $d_0, d_1,\ldots, d_{2N-2}$ are information words.

\subsection{Specific examples of matrix generators of $Z_N$ and the corresponding RAID-6 systems.\label{26}}
Now we are ready to construct explicit examples of RAID-6 based
on the theory of cones developed in the above subsections. The non-extended code, based on the Sylvester matrix, is known as EVENODD code \cite{bbbm}.

{\bf Lemma \ref{26}.1.} {\it Let $S_{N}$ be the $(N-1)\times
(N-1)$ Sylvester matrix,
\begin{displaymath}
 S_{N}=\left( \begin{array}{ccccccc}
 0 & 0&0&\cdot&\cdot&\cdot&1 \\
 1 & 0&0&0&\cdot&\cdot&1 \\
 0 & 1&0&0&0&\cdot&1 \\
 \cdot & \cdot&\cdot&\cdot&\cdot&\cdot&\cdot \\
 0 & 0&0&\cdot&1&0&1 \\
 0 & 0&0&\cdot&\cdot&1&1 \\
 \end{array} \right).
 \end{displaymath}
Then
\begin{mylist}
\item[(i)] $S_{N}$ has order $N$.
\item[(ii)] Matrix $Id+S_{N}$ is non-degenerate if $N$ is
odd and is degenerate if $N$ is even.
\end{mylist} }
{\bf Proof. (i)} An explicit computation shows, that for any
$(N-1)$-dimensional binary vector $x$ and for any $1 \leq k \leq
N$,
 \bea
 S_N^k \left( \begin{array}{c}
 x_{N-1}\\
 x_{N-2}\\
 x_{N-3}\\
 \cdot\\
 \cdot\\
 \cdot\\
 \cdot\\
 x_1
 \end{array} \right)
=\left( \begin{array}{c}
 x_k\\
 x_{k}\\
 x_{k}\\
 \cdot\\
 \cdot\\
 \cdot\\
 \cdot\\
 x_k \end{array} \right) +
 \left( \begin{array}{c}
 x_{k-1}\\
 x_{k-2}\\
 \cdot\\
 x_{1}\\
 0\\
 x_{N-1}\\
 \cdot\\
 x_{k+1}
\end{array} \right).\label{powers}
\eea
 In the above formula $x_j \equiv 0$, unless $1\leq j \leq (N-1)$.
 Therefore, $S_N^k \neq Id$, for any $1\leq k \leq N-1$.
 Setting $k=N$ in the above formula, we get $S_{N}^{N}x=x$ for
 any $x$, which implies that $S_N^{N}=Id$. Therefore, the order of the matrix $S_N$ is $N$.

 {\bf (ii)} The characteristic polynomial of $S_N$ is
$f(x)=\sum_{k=0}^{N-1}x^k$. (In order to prove this it is
sufficient to notice that the matrix $S_N$ is the companion matrix
of the polynomial $f(x)$ \cite{lidl}. As such, $f(x)$ is both the
characteristic and the minimal polynomial of the matrix $S_N$.)
Therefore,
$$
f(S_N)=\sum_{k=0}^{N-1}S_{N}^k=0.
$$
 Notice that the matrix $S_N$ is non-degenerate as it has a
 positive order.
If $N$ is odd, we can re-write the characteristic polynomial as
 $$
 f(S_N)=(Id+S_{N})(1+S_N+S_N^3+\ldots+S_N^{(N-3)})+S_N^{N-1}
 $$
 Therefore, the degeneracy of $Id+S_N$ will contradict the non-degeneracy
 of $S_N$. If $N$ is even, the sum of all rows of $Id+S_N$ is
 zero, which implies degeneracy.  {\bf QED}

 Lemma  \ref{26}.1 states that the matrix $S_N$ generates the cyclic group
 $Z_N$ and that the matrix $Id+S_N$ is non-degenerate for any odd $N$.
 Given that $N$ is an odd prime, Corollary \ref{24}.4 implies that
 using parity equations (\ref{znparity}) with $g=S_N$, it is
 possible to protect $N$ data disks against the failure of any two
 disks. Furthermore, if $N>3$ is a Fermat prime or 2 is a primitive root modulo $N$, $2N-1$ data disks can
 be protected against double failure thanks to the results of section \ref{23}.

 We will refer to the RAID-6 system based on Sylvester matrix $S_N$ as
 $Z_N$-RAID. Let us give several examples of such systems.
\begin{mylist}
\item{(1)} $Z_3$-RAID has been considered in subsections \ref{21},
\ref{22}. It can protect up to $3$ information disks against double failure.
As $N=3$, protection of $5$ information disks using extended $Q$-parity (\ref{qext}) is impossible.
\item{(2)} Using $Z_{17}$-RAID, one can protect up to $N=17$ disks using
$Q$-parity (\ref{znparity}) and up to $2N-1=33$ disks using
extended $Q$-parity (\ref{qext}). 
\item{(3)} Using $Z_{257}$-RAID, one can protect up to $N=257$ disks
using $Q$-parity  (\ref{znparity}) and up to $2N-1=513$ disks
using extended $Q$-parity (\ref{qext}). 
\end{mylist}

 It can be seen from (\ref{powers}), that the multiplication of data vectors with any power of
 the Sylvester matrix $S_N$ requires one left and one right shift, one $n$-bit XOR and
 one AND only. Thus the operations of updating $Q$-parity and recovering data within $Z_N$-RAID
 does not require any special instructions, such as Galois
 field look-up tables for logarithms and products. As a
 result, the implementation of $Z_N$-RAID can in some cases be more efficient
 and quick than the implementation of the more conventional Reed-Solomon based
 RAID-6. In the next section we will demonstrate the advantage of
 $Z_N$-RAID using an example of Linux kernel implementation of
 $Z_{17}$-RAID system.

\section{Linux Kernel $Z_N$-RAID Implementation}

\subsection{Syndrome Calculation for the Reed-Solomon RAID-6.}
\label{parityeq}
  First, let us briefly recall the RAID-6 scheme based on Reed-Solomon code in
  the Galois field $\tF$, see \cite{peter} for more details.
  Let $D_0, \ldots, D_{N-1}$ be the bytes of data from $N$ information
  disks. Then the parity bytes $P$ and $Q$ are computed as
  follows, 
  using \footnote{
Algebraically, we use the standard representation in electronics:
$\tF=GF(2)[x]/I$ where the ideal $I$ is generated by
$x^8+x^4+x^3+x^2+1$ and $g=x+I$
}
$g = \{02\} \in \tF$:
  \begin{ea}
    P &= D_0 + D_1 + \ldots + D_{N-1} , \ 
    Q &= D_0 + g D_1 + \ldots + g^{N-1} D_{N-1}.\label{rsparity}
  \end{ea}

\begin{flushleft}
  The multiplication by $g = \{02\}$ can be viewed as the following matrix multiplication.
  \begin{ea}
    \begin{bmatrix}
      y_0 \\ y_1 \\ y_2 \\ y_3 \\ y_4 \\ y_5 \\ y_6 \\ y_7
    \end{bmatrix}
    =
    \begin{bmatrix}
      0 & 0 & 0 & 0 & 0 & 0 & 0 & 1 \\
      1 & 0 & 0 & 0 & 0 & 0 & 0 & 0 \\
      0 & 1 & 0 & 0 & 0 & 0 & 0 & 1 \\
      0 & 0 & 1 & 0 & 0 & 0 & 0 & 1 \\
      0 & 0 & 0 & 1 & 0 & 0 & 0 & 1 \\
      0 & 0 & 0 & 0 & 1 & 0 & 0 & 0 \\
      0 & 0 & 0 & 0 & 0 & 1 & 0 & 0 \\
      0 & 0 & 0 & 0 & 0 & 0 & 1 & 0 \\
    \end{bmatrix}
    \begin{bmatrix}
      x_0 \\ x_1 \\ x_2 \\ x_3 \\ x_4 \\ x_5 \\ x_6 \\ x_7
    \end{bmatrix}
    =
    \begin{bmatrix}
      x_7 \\ x_0 \\ x_1 \xor x_7 \\ x_2 \xor x_7 \\ x_3 \xor x_7 \\ x_4 \\ x_5 \\ x_6
    \end{bmatrix}\label{g02}
  \end{ea}
Given (\ref{g02}), parity equations (\ref{rsparity}) become similar to (\ref{znparity}).
Indeed, the element $g$ generates a cyclic group, so a 2-error correcting Reed-Solomon code is a partial case of a cone based RAID.
However, $Z_N$-RAID has several advantages. 
For instance, using Sylvester matrices one can
achieve a simpler implementation of matrix multiplication.
\end{flushleft}

\subsection{Linux Kernel Implementation of Syndrome Calculation}
\label{unrolledq}
  To compute the $Q$-parity, we rewrite (\ref{rsparity}) as
  \begin{ea}
    Q = D_0 + g (D_1 + g (\ldots + g (D_{N-2} + g D_{N-1})\ldots)))
  \end{ea}
which requires $(N-1)$ multiplications by $g = \{02\}$.

\begin{flushleft}
  The product $y$ of a single byte $x$ and $g=\{02\}$  can be implemented as
  follows.
\end{flushleft}
\begin{center}
  \texttt{uint8\_t\ x,\ y;} \hspace{20mm}
  \texttt{y\ =\ (x\ <<\ 1)\ $\hat{}$\ ((x\ \&\ 0x80)\ ?\ 0x1d\ :\ 0x00);} 
\end{center}
\begin{flushleft}
  Notice that \texttt{(x\ \&\ 0x80)} picks out $x_7$ from $x$, so
$$
  \texttt{((x\ \&\ 0x80)\ ?\ 0x1d\ :\ 0x00)}
$$
 selects between the two bit patterns
  $00011101$ and $00000000$ depending on $x_7$.  
Since the carry is discarded
  from \texttt{(x\ <<\ 1)},  
\begin{ea}    
\text{\texttt{(x\ <<\ 1)}}  =[x_6, x_5, x_4, x_3, x_2, x_1, x_0, 0] 
    \text{\texttt{((x\ \&\ 0x80)\ ?\ 0x1d\ :\ 0x00)}}
=[0, 0, 0, x_7, x_7, x_7, 0, x_7].
  \end{ea}
  We can also implement the multiplication as follows.
\end{flushleft}
\begin{center}
  \texttt{int8\_t\ x,\ y;} \hspace{20mm}
  \texttt{y\ =\ (x\ +\ x)\ $\hat{}$\ (((x\ <\ 0)\ ?\ 0xff\ :\ 0x00)\ \&\ 0x1d);}
\end{center}
\begin{flushleft}
  Here we treat the values as signed, rather than unsigned.  Whilst this implementation appears
  more complex than the first (since it uses addition and comparison), it can efficiently be
  implemented using SIMD instructions on modern processors, such as MMX/SSE/SSE2/AltiVec.
\end{flushleft}
\begin{flushleft}
  In particular, we will use the following four SSE2 instructions, which store the result in place of the second operand:
\end{flushleft}
\begin{flushleft}
  \texttt{pxor\ x,\ y\ \ \ \ \ \ :\ y\ =\ x\ $\hat{}$\ y;} \hspace{25mm}
  \texttt{pand\ x,\ y\ \ \ \ \ \ :\ y\ =\ x\ \&\ y;} \\
  \texttt{paddb\ x,\ y\ \ \ \ \ :\ y\ =\ x\ +\ y;} \hspace{23mm}
  \texttt{pcmpgtb\ x,\ y\ \ \ :\ y\ =\ (y > x)\ ?\ 0xff\ :\ 0x00;} \\
\end{flushleft}
\begin{flushleft}
  Therefore we can implement a single multiplication with the following pseudo SSE2 assembler code. We assume that the variables $\texttt{y}$ and $\texttt{c}$ are initialised as \texttt{y\ =\ 0} and \texttt{c\ =\ 0x1d}.
\end{flushleft}
\begin{flushleft}
  \texttt{pcmpgtb\ x,\ y\ \ \ :\ y\ =\ (x < 0)\ ?\ 0xff\ :\ 0x00;\ //\ (x < 0)\ ?\ 0xff\ :\ 0x00} \\
  \texttt{paddb\ x,\ x\ \ \ \ \ :\ x\ =\ x\ +\ x;\ \ \ \ \ \ \ \ \ \ \ \ \ \ \ \ \ //\ x\ +\ x} \\
  \texttt{pand\ c,\ y\ \ \ \ \ \ :\ y\ =\ y\ \&\ 0x1d;\ \ \ \ \ \ \ \ \ \ \ \ \ \ //\ ((x < 0)\ ?\ 0xff\ :\ 0x00)\ \&\ 0x1d} \\
  \texttt{pxor\ x,\ y\ \ \ \ \ \ :\ y\ =\ x\ $\hat{}$\ y;\ \ \ \ \ \ \ \ \ \ \ \ \ \ \ \ \ \ //\ (x\ +\ x)\ $\hat{}$} \\
  \texttt{\ \ \ \ \ \ \ \ \ \ \ \ \ \ \ \ \ \ \ \ \ \ \ \ \ \ \ \ \ \ \ \ \ \ \ \ \ \ \ \ \ \ \ \ //\ \ (((x < 0)\ ?\ 0xff\ :\ 0x00)\ \&\ 0x1d)} \\
\end{flushleft}
\begin{flushleft}
The comparison operation overwrites the constant 0 stored in \texttt{y}. Therefore, when we implement
  the complete algorithm we must recreate the constant before each multiplication.  We can do it as follows.
\end{flushleft}
\begin{flushleft}
  \texttt{pxor\ y,\ y\ \ \ \ \ \ :\ y\ =\ y\ $\hat{}$\ y;\ \ \ \ \ \ \ \ \ \ \ \ \ \ \ \ \ \ //\ y\ $\hat{}$\ y\ =\ 0} \\
\end{flushleft}
\begin{flushleft}
Besides the five instruction above we need three 
other instructions
to complete the inner loop of the algorithm.
They are multiply, fetch a new byte of data $D$ and update the parity variables $P$ and $Q$:
  \begin{ea}
    P &= D + P, \hspace{20mm}
    Q &= D + g Q.
  \end{ea}
The complete algorithm requires the following eight instructions.
\end{flushleft}
\begin{flushleft}
  \texttt{pxor\ y,\ y\ \ \ \ \ \ :\ y\ =\ y\ $\hat{}$\ y;\ \ \ \ \ \ \ \ \ \ \ \ \ \ \ \ \ \ //\ y\ $\hat{}$\ y\ =\ 0} \\
  \texttt{pcmpgtb\ q,\ y\ \ \ :\ y\ =\ (q < 0)\ ?\ 0xff\ :\ 0x00;\ //\ (q < 0)\ ?\ 0xff\ :\ 0x00} \\
  \texttt{paddb\ q,\ q\ \ \ \ \ :\ q\ =\ q\ +\ q;\ \ \ \ \ \ \ \ \ \ \ \ \ \ \ \ \ //\ q\ +\ q} \\
  \texttt{pand\ c,\ y\ \ \ \ \ \ :\ y\ =\ y\ \&\ 0x1d;\ \ \ \ \ \ \ \ \ \ \ \ \ \ //\ ((q < 0)\ ?\ 0xff\ :\ 0x00)\ \&\ 0x1d} \\
  \texttt{pxor\ y,\ q\ \ \ \ \ \ :\ q\ =\ q\ $\hat{}$\ y;\ \ \ \ \ \ \ \ \ \ \ \ \ \ \ \ \ \ //\ g.q\ =\ (q\ +\ q)\ $\hat{}$} \\
  \texttt{\ \ \ \ \ \ \ \ \ \ \ \ \ \ \ \ \ \ \ \ \ \ \ \ \ \ \ \ \ \ \ \ \ \ \ \ \ \ \ \ \ \ \ \ //\ \ (((q < 0)\ ?\ 0xff\ :\ 0x00)\ \&\ 0x1d)} \\
  \texttt{movdqa d[i], d\ :\ d\ =\ d[i]\ \ \ \ \ \ \ \ \ \ \ \ \ \ \ \ \ \ \ //\ d[i]} \\
  \texttt{pxor\ d,\ q\ \ \ \ \ \ :\ q\ =\ d\ $\hat{}$\ q;\ \ \ \ \ \ \ \ \ \ \ \ \ \ \ \ \ \ //\ d[i]\ $\hat{}$\ p} \\
  \texttt{pxor\ d,\ p\ \ \ \ \ \ :\ p\ =\ d\ $\hat{}$\ p;\ \ \ \ \ \ \ \ \ \ \ \ \ \ \ \ \ \ //\ d[i]\ $\hat{}$\ g.q} \\
\end{flushleft}

\begin{flushleft}
We can gain a further increase in speed by partially unrolling the 'for' loop around the inner loop.
\end{flushleft}

\subsection{Reconstruction}
We consider a situation that
two data disks $D_x$ and $D_y$ have failed.  We must reconstruct $D_x$ and $D_y$ from
the remaining data disks $D_i$ $(i \ne x,y)$ and the parity disks $P$ and $Q$, see (\ref{rsparity}).
Let us define $P_{xy}$ and $Q_{xy}$ as the syndromes under an assumption
that the failed disks were zero: 
  \begin{ea} \label{PQxy}
    P_{xy} = \sum_{i \ne x,y}{D_i}, \hspace{20mm}
    Q_{xy} = \sum_{i \ne x,y}{g^i D_i}.
  \end{ea}
Rewriting (\ref{rsparity}) in the light of (\ref{PQxy}),
  \begin{ea}\label{reconstructfromparity}
    D_x + D_y = P + P_{xy}, \hspace{20mm}
    g^x D_x + g^y D_y = Q + Q_{xy}. 
  \end{ea}
Let us define
  \begin{ea}
    A = (1 + g^{y-x})^{-1}, \hspace{20mm}
    B = g^{-x} (1 + g^{y-x})^{-1}.
  \end{ea}
Now we  eliminate $D_x$ from equations (\ref{reconstructfromparity}):
  \begin{ea}
    D_y = (1 + g^{y-x})^{-1} (P + P_{xy}) +  g^{-x} (1 + g^{y-x})^{-1} (Q + Q_{xy}) = A (P + P_{xy}) + B (Q + Q_{xy}).
  \end{ea}
Finally, $D_x$ is computed from $D_y$ by the back substitution into \eqref{reconstructfromparity}:
  \begin{ea}
    D_x &= D_y + (P + P_{xy}).
  \end{ea}

\subsection{Linux Kernel Implementation of Reconstruction}
  We compute the following values in $\tF$:
  \begin{ea}
    A   &= (1 + g^{y-x})^{-1}, &\hspace{20mm}&
    B   &= g^{-x} (1 + g^{y-x})^{-1} &= (g^x + g^y)^{-1}, \\
    D_y &= A (P + P_{xy}) + B (Q + Q_{xy}), &\hspace{20mm}&
    D_x &= D_y + (P + P_{xy}). &
  \end{ea}
It is worth pointing out 
that for specific $x$ and $y$, we only need to compute $A$ and $B$ once.
  The Linux kernel provides the following look-up tables:
\begin{flushleft}
  \texttt{raid6\_gfmul[256][256]\ \ :\ $xy$} \hspace{25mm}
  \texttt{raid6\_gfexp[256]\ \ \ \ \ \ \ :\ $g^x$} \\
  \texttt{raid6\_gfinv[256]\ \ \ \ \ \ \ :\ $x^{-1}$} \hspace{22mm}
  \texttt{raid6\_gfexi[256]\ \ \ \ \ \ \ :\ $(1+g^x)^{-1}$} \\
\end{flushleft}
\begin{flushleft}
Using this, we compute $A$ and $B$ as follows:
\end{flushleft}
\begin{flushleft}
  \texttt{A\ =\ raid6\_gfexi[y-x]} \hspace{12mm} and \hspace{10mm}
  \texttt{B\ =\ raid6\_gfinv[raid6\_gfexp[x]\ $\hat{}$\ raid6\_gfexp[y]]} \\
\end{flushleft}
\begin{flushleft}
To reconstruct $D_x$ and $D_y$ we start by 
constructing $P_{xy}$ and $Q_{xy}$ using the standard syndrome code.
Then we execute the following code.
\end{flushleft}
\begin{flushleft}
\texttt{dP\ =\ P\ $\hat{}$\ Pxy;
\hspace{12mm}
//\ $P + P_{xy}$}
\newline
\texttt{dQ\ =\ Q\ $\hat{}$\ Qxy;
\hspace{12mm}
//\ $Q + Q_{xy}$}
\newline
\texttt{Dy\ =\ raid6\_gfmul[A][dP]\ $\hat{}$\ raid6\_gfmul[B][dQ];
\hspace{12mm}
//\ $A (P + P_{xy}) + B (Q + Q_{xy})$}
\newline
\texttt{Dx\ =\ Dy\ $\hat{}$\ dP;
\hspace{12mm}
//\ $D_y + (P + P_{xy})$} \\
\end{flushleft}

\subsection{$Z_{17}$-RAID Implementation}
The multiplication by the Sylvester matrix $g$ looks like
  \begin{ea}
    \begin{bmatrix}
      y_0 \\ y_1 \\ y_2 \\ y_3 \\ y_4 \\ y_5 \\ y_6 \\ y_7 \\ y_8 \\ y_9 \\ y_{10} \\ y_{11} \\ y_{12} \\ y_{13} \\ y_{14} \\ y_{15}
    \end{bmatrix}
=    
\begin{bmatrix}
      0 & 0 & 0 & 0 & 0 & 0 & 0 & 0 & 0 & 0 & 0 & 0 & 0 & 0 & 0 & 1 \\
      1 & 0 & 0 & 0 & 0 & 0 & 0 & 0 & 0 & 0 & 0 & 0 & 0 & 0 & 0 & 1 \\
      0 & 1 & 0 & 0 & 0 & 0 & 0 & 0 & 0 & 0 & 0 & 0 & 0 & 0 & 0 & 1 \\
      0 & 0 & 1 & 0 & 0 & 0 & 0 & 0 & 0 & 0 & 0 & 0 & 0 & 0 & 0 & 1 \\
      0 & 0 & 0 & 1 & 0 & 0 & 0 & 0 & 0 & 0 & 0 & 0 & 0 & 0 & 0 & 1 \\
      0 & 0 & 0 & 0 & 1 & 0 & 0 & 0 & 0 & 0 & 0 & 0 & 0 & 0 & 0 & 1 \\
      0 & 0 & 0 & 0 & 0 & 1 & 0 & 0 & 0 & 0 & 0 & 0 & 0 & 0 & 0 & 1 \\
      0 & 0 & 0 & 0 & 0 & 0 & 1 & 0 & 0 & 0 & 0 & 0 & 0 & 0 & 0 & 1 \\
      0 & 0 & 0 & 0 & 0 & 0 & 0 & 1 & 0 & 0 & 0 & 0 & 0 & 0 & 0 & 1 \\
      0 & 0 & 0 & 0 & 0 & 0 & 0 & 0 & 1 & 0 & 0 & 0 & 0 & 0 & 0 & 1 \\
      0 & 0 & 0 & 0 & 0 & 0 & 0 & 0 & 0 & 1 & 0 & 0 & 0 & 0 & 0 & 1 \\
      0 & 0 & 0 & 0 & 0 & 0 & 0 & 0 & 0 & 0 & 1 & 0 & 0 & 0 & 0 & 1 \\
      0 & 0 & 0 & 0 & 0 & 0 & 0 & 0 & 0 & 0 & 0 & 1 & 0 & 0 & 0 & 1 \\
      0 & 0 & 0 & 0 & 0 & 0 & 0 & 0 & 0 & 0 & 0 & 0 & 1 & 0 & 0 & 1 \\
      0 & 0 & 0 & 0 & 0 & 0 & 0 & 0 & 0 & 0 & 0 & 0 & 0 & 1 & 0 & 1 \\
      0 & 0 & 0 & 0 & 0 & 0 & 0 & 0 & 0 & 0 & 0 & 0 & 0 & 0 & 1 & 1 \\
    \end{bmatrix}
    \begin{bmatrix}
      x_0 \\ x_1 \\ x_2 \\ x_3 \\ x_4 \\ x_5 \\ x_6 \\ x_7 \\ x_8 \\ x_9 \\ x_{10} \\ x_{11} \\ x_{12} \\ x_{13} \\ x_{14} \\ x_{15}
    \end{bmatrix}
    =
    \begin{bmatrix}
      x_{15} \\ x_0 \xor x_{15} \\ x_1 \xor x_{15} \\ x_2 \xor x_{15} \\ x_3 \xor x_{15} \\ x_4 \xor x_{15} \\ x_5 \xor x_{15} \\ x_6 \xor x_{15} \\
      x_7 \xor x_{15} \\ x_8 \xor x_{15} \\ x_9 \xor x_{15} \\ x_{10} \xor x_{15} \\ x_{11} \xor x_{15} \\ x_{12} \xor x_{15} \\ x_{13} \xor x_{15} \\ x_{14} \xor x_{15}
    \end{bmatrix} .
  \end{ea} 
\begin{flushleft}
We implement the multiplication of a double byte $y = g x$ as follows:
\end{flushleft}
\begin{flushleft}
  \texttt{int16\_t\ x,\ y;} \hspace{15mm}
  \texttt{y\ =\ (x\ +\ x)\ $\hat{}$\ ((x\ <\ 0)\ ?\ 0xffff\ :\ 0x0000);}
\end{flushleft}
\begin{flushleft}
  We can implement this in assembler using the following seven
  instructions.
\end{flushleft}
\begin{flushleft}
  \texttt{pxor\ y,\ y\ \ \ \ \ \ :\ y\ =\ y\ $\hat{}$\ y;\ \ \ \ \ \ \ \ \ \ \ \ \ \ \ \ \ \ \ \ \ \ //\ y\ $\hat{}$\ y\ =\ 0} \\
  \texttt{pcmpgtw\ q,\ y\ \ \ :\ y\ =\ (q < 0)\ ?\ 0xffff\ :\ 0x0000;\ //\ (q < 0)\ ?\ 0xffff\ :\ 0x0000} \\
  \texttt{paddw\ q,\ q\ \ \ \ \ :\ q\ =\ q\ +\ q;\ \ \ \ \ \ \ \ \ \ \ \ \ \ \ \ \ \ \ \ \ //\ q\ +\ q} \\
  \texttt{pxor\ y,\ q\ \ \ \ \ \ :\ q\ =\ q\ $\hat{}$\ y;\ \ \ \ \ \ \ \ \ \ \ \ \ \ \ \ \ \ \ \ \ \ //\ g.q\ =\ (q\ +\ q)\ $\hat{}$} \\
  \texttt{\ \ \ \ \ \ \ \ \ \ \ \ \ \ \ \ \ \ \ \ \ \ \ \ \ \ \ \ \ \ \ \ \ \ \ \ \ \ \ \ \ \ \ \ \ \ \ \ //\ \ ((q < 0)\ ?\ 0xffff\ :\ 0x0000)} \\
  \texttt{movdqa d[i], d\ :\ d\ =\ d[i]\ \ \ \ \ \ \ \ \ \ \ \ \ \ \ \ \ \ \ \ \ \ \ //\ d[i]} \\
  \texttt{pxor\ d,\ q\ \ \ \ \ \ :\ q\ =\ d\ $\hat{}$\ q;\ \ \ \ \ \ \ \ \ \ \ \ \ \ \ \ \ \ \ \ \ \ //\ d[i]\ $\hat{}$\ p} \\
  \texttt{pxor\ d,\ p\ \ \ \ \ \ :\ p\ =\ d\ $\hat{}$\ p;\ \ \ \ \ \ \ \ \ \ \ \ \ \ \ \ \ \ \ \ \ \ //\ d[i]\ $\hat{}$\ g.q} \\
\end{flushleft}

\begin{flushleft}
  Below are the results of the Linux kernel RAID-6  algorithm selection programs,  aimed to select the fastest implementation of the algorithm.
  Algorithms using CPU/MMX/SSE/SSE2 instructions with various levels of unrolling are compared.  
The results were obtained from a 2.8 GHz Intel Pentium 4 (x86).
\end{flushleft}
\begin{flushleft}
  \texttt{\hspace{37mm} DP-RAID \hspace{26mm} $Z_{17}$-RAID} \\
  \texttt{int32x1 \hspace{22mm}  694\ MB/s \hspace{25mm} 766\ MB/s} \\
  \texttt{int32x2 \hspace{22mm}   939\ MB/s \hspace{25mm} 854\ MB/s} \\
  \texttt{int32x4 \hspace{22mm}   635\ MB/s \hspace{25mm} 838\ MB/s} \\
  \texttt{int32x8 \hspace{22mm} 505\ MB/s \hspace{25mm} 604\ MB/s} \\
  \texttt{mmxx1  \hspace{25mm} 1893\ MB/s \hspace{24mm} 2117\ MB/s} \\
  \texttt{mmxx2  \hspace{25mm} 2025\ MB/s \hspace{24mm} 2301\ MB/s} \\
  \texttt{sse1x1 \hspace{23mm} 1200\ MB/s \hspace{24mm} 1284\ MB/s} \\
  \texttt{sse1x2 \hspace{23mm} 2000\ MB/s \hspace{24mm} 2263\ MB/s} \\
  \texttt{sse2x1 \hspace{23mm} 1850\ MB/s \hspace{24mm} 2357\ MB/s} \\
  \texttt{sse2x2 \hspace{23mm} 2702\ MB/s \hspace{24mm}\ 3160\ MB/s} \\
\end{flushleft}
\begin{flushleft}
  Comparing the above results against the standard Linux kernel results shows an average of $14.5\%$ speed increase and
  an increase of $16.9\%$ for the fastest sse2x2 implementation.  This is consistent with the theoretical increase of $14.3\%$
  for seven instructions instead of eight instructions. It is worth mentioning
  that no look-up tables have been used to implement $Z_{17}$-RAID.
\end{flushleft}

\subsection{$Z_N$ RAID Reconstruction}
  We need to compute the following matrices and vectors:
  \begin{ea}
    A   &= (1 + g^{y-x})^{-1}, &\hspace{12mm} &
    B   &= g^{-x} (1 + g^{y-x})^{-1} &= (g^x + g^y)^{-1} \\
    D_y &= A (P + P_{xy}) + B (Q + Q_{xy}), &\hspace{10mm} &
    D_x &= D_y + (P + P_{xy}) . &
  \end{ea}
  We rewrite them as follows:
  \begin{ea}
    z   &= y - x,  & \Delta P &= P + P_{xy}, & 
    \Delta Q &= Q + Q_{xy} \\
    D_y &= (1 + g^z)^{-1} \Delta P & + g^{-x} (1 + g^z)^{-1} \Delta Q, 
& &D_x = D_y + \Delta P.  \\
  \end{ea}
Using the standard identities
$g^{-k} = g^{17-k}$
and 
$(1 + g)^{-1} = 1 + g^2 + \ldots + g^{16}$,
we derive new identities:
  \begin{ea}
    (1 + g^z)^{-1} = 1 + g^{2z} + g^{4z} + \ldots + g^{16z} ,  \hspace{43mm} \\
    g^{-x} (1 + g^z)^{-1} = g^{17-x} (1 + g^z)^{-1} = g^{17-x} (1 + g^{2z} + g^{4z} + \ldots + g^{16z}). 
  \end{ea}
Consequently, we need to compute
  \begin{ea} \label{delP}
    (1 + g^z)^{-1} \Delta P = (1 + g^{2z} + g^{4z} + \ldots + g^{16z}) \Delta P = \hspace{38mm} \\
    = 1 + g^{2z} (1 + g^{2z} (1 + g^{2z} (1 + g^{2z} (1 +g^{2z} (1 + g^{2z} (1 + g^{2z} (1 + g^{2z} \Delta P))))))) \\
  \end{ea}
  and
  \begin{ea} \label{delQ}
    g^{-x} (1 + g^z)^{-1} \Delta Q
    = g^{17-x} (1 + g^{2z} + g^{4z} + \ldots + g^{16z}) \Delta Q = \\
    = g^{17-x} (1 + g^{2z} (1 + g^{2z} (1 + g^{2z} (1 + g^{2z} (1 + & g^{2z} (1 + g^{2z} (1 + g^{2z} (1 + g^{2z} \Delta Q)))))))) \\
  \end{ea}
Both \eqref{delP} and \eqref{delQ}
require only one principle operation, multiplication by $g^k$.
\begin{flushleft}
The multiplication of a single word $y = g^k x$ for $1 \leq k \leq 16$ can be implemented as follows.
\end{flushleft}
\begin{flushleft}
  \texttt{int16\_t\ x,\ y;} \\
  \texttt{y\ =\ (x\ <<\ k)\ $\hat{}$\ (x\ >>\ (17\ -\ k))\ $\hat{}$\ (((x\ <<\ (k\ -\ 1))\ <\ 0)\ ?\ 0xffff\ :\ 0x0000);}
\end{flushleft}
\begin{flushleft}
We precompute $m=k-1$ and $n=17-k$ as they remain constant during reconstruction.
This leads to the following assembler implementation.
\end{flushleft}
\begin{flushleft}
  \texttt{pxor\ y,\ y\ \ \ \ :\ y\ =\ 0} \hspace{60mm}
  \texttt{movdqa\ x,\ z\ \ :\ z\ =\ x} \\
  \texttt{psllw\ m,\ z\ \ \ :\ z\ =\ x\ <<\ (k-1)} \\
  \texttt{pcmpgtw\ z,\ y\ :\ y\ =\ (((x\ <<\ (k-1))\ <\ 0)\ ?\ 0xffff\ :\ 0x0000} \\
  \texttt{paddw\ z,\ z\ \ \ :\ z\ =\ x\ <<\ k} \\
  \texttt{pxor\ z,\ y\ \ \ \ :\ y\ =\ (((x\ <<\ (k-1))\ <\ 0)\ ?\ 0xffff\ :\ 0x0000)\ $\hat{}$\ (x\ <<\ k)} \\
  \texttt{psrlw\ n,\ x\ \ \ :\ x\ =\ x\ >>\ (17-k)} \hspace{42mm}
  \texttt{pxor\ x,\ y\ \ \ \ :\ y\ =\ g\ $\hat{}$\ k\ x} \\
\end{flushleft}
\begin{flushleft}
  Below is a table showing benchmark results of complete reconstruction algorithm implemented using SSE2 assembler
  and the standard Linux kernel look-up table reconstruction implementation, for the cases of double data disk failure,
  double disk failure of one data disk and the P-parity disk, and double parity disk failure.  Note the data represents
  time taken to complete benchmark, so lower is better.
\end{flushleft}
\begin{center}
  \begin{tabular}{c|c c c}
    Failure      &   DD &   DP &  PQ \\ \hline
    DP-RAID & 2917 & 2771 & 905 \\
    $Z_{17}$-RAID       & 2711 & 1274 & 809 \\
  \end{tabular}
\end{center}
\begin{flushleft}
  Comparing the complete reconstruction algorithm implemented using SSE2 assembler against the standard Linux kernel 
  look-up table implementation, shows approximately $7\%$ speed increase for $DD$ failure, $54\%$ speed increase for $DP$ failure 
  and $11\%$ speed increase for $PQ$ failure.
\end{flushleft}

\section{Conclusions.}
In this paper we have demonstrated that {\em cones}
provide a natural framework for the design of RAID.
They provide a flexible approach that can be used to design
a system.
It is worth further theoretical
investigation what other examples of cones can be constructed or
what the maximal possible size of a cone is.

We have also demonstrated that cyclic groups give rise
to natural and convenient to operate examples of cones.
One particular advantage is that $Z_N$-RAID does not
require support of the Galois field operations. 

On the practical side, $Z_{17}$-RAID and $Z_{257}$-RAID 
are breakthrough techniques that show at least 10\% 
improvement during simulations compared to DP-RAID.

%


\section*{Acknowledgment}
The authors would like to thank Arithmatica, Ltd. for the opportunity
to use its research facilities. The authors would also like
to thank Robert Maddock and Igor Shparlinski for valuable information 
on the subject of the paper. 
Finally, the authors are indebted to Samir Siksek for the interest in the prime number condition
that appears in this paper and computation of several primes satisfying it.



%

\end{document}